# Super-resolution imaging of quantum emitters in layered materials

Mehran Kianinia [#, †], Carlo Bradac [#,†,*], Fan Wang[†], Bernd Sontheimer[‡], Toan Trong Tran[†], Minh Nguyen[†], Sejeong Kim[†], Zai-Quan Xu[†], Dayong Jin[†], Andreas W. Schell[§], Charlene J. Lobo[†], Igor Aharonovich[†, *], and Milos Toth[†, *]

[†] School of Mathematical and Physical Sciences, University of Technology Sydney, Ultimo, NSW, 2007, Australia.

[‡] Institut für Physik, Humboldt-Universität zu Berlin, 12489 Berlin, Germany.

[§] Department of Electronic Science and Engineering, Kyoto University, 615-8510 Kyoto, Japan.

[#] contributed equally

*correspondence to Carlo.Bradac@uts.edu.au, igor.aharonovich@uts.edu.au
milos.toth@uts.edu.au

**Layered van der Waals materials are emerging as compelling two-dimensional (2D) platforms for studies of nanophotonics, polaritonics, valleytronics and spintronics, and have the potential to transform applications in sensing, imaging and quantum information processing. Amongst these, hexagonal boron nitride (hBN) is unique in that it hosts ultra-bright, room temperature single photon emitters (SPEs). However, an outstanding challenge is to locate SPEs in hBN with high precision – a task which requires breaking the optical diffraction limit. Here, we report the imaging of SPEs in layered hBN with a spatial resolution of (63±4) nm using ground state depletion (GSD) nanoscopy. Furthermore, we show that SPEs in hBN possess nonlinear photophysical properties which can be used to realize a new variant of GSD that employs a coincident pair of doughnut-shaped lasers to reduce the laser power that is needed to achieve a given resolution target. Our findings expand the current understanding of the photophysics of quantum emitters in layered hBN and demonstrate the potential for advanced nanophotonic and bio-imaging applications which require localization of individual emitters with super-resolution accuracy.**

The exploration of nanophotonic phenomena in 2D systems using materials such as transition metal dichalcogenides (TMDCs), phosphorene and hexagonal boron nitride (hBN) has been gaining momentum in recent years[1-8]. Localized nanoscale effects, including radiative decay of interlayer excitons and emission of anti-bunched photons from deep trap-point defects, are particularly interesting and important[9-11]. The latter, for instance, is key to the practical deployment of scalable, on-chip quantum photonic devices[12, 13]. In this context, 2D-layered hBN has shown great promise owing to its ability to host fully polarized, ultra-bright color centres which act as SPEs at – and beyond – room temperature[9, 14-18]. The current challenge towards the realization of practical, scalable, integrated quantum photonic circuits, is the efficient coupling of the emitters to optical and mechanical resonators. The ability to locate and ultimately fabricate SPEs with a degree of precision well beyond the optical diffraction limit of visible light is thus becoming imperative. Similarly, full exploitation of the emitters for sensing and biological imaging applications requires that the SPEs be compatible with super-resolution imaging techniques.

Super-resolution imaging has so far been realized using fluorophores that include single molecules, color centers and quantum dots [19-21]. Techniques like photo-activated localization microscopy (PALM)[22] and stochastic optical reconstruction microscopy (STORM)[23] rely on ultra-fast blinking of fluorophores, and require the acquisition of multiple image frames and resource-intensive data analysis and post-processing steps. Nanoscopy techniques such as stimulated emission depletion (STED)[24] and ground state depletion (GSD)[25], on the other hand, enable direct imaging with spatial resolutions beyond 10 nm, and remove the need for post-acquisition image processing. However, STED and GSD of individual solid-state quantum emitters (i.e. point defects in solids) have so far only been realized with the negatively charged nitrogen vacancy (NV$^-$) center in diamond[19, 26]. This is because both STED and GSD impose a stringent set of criteria on the properties of the emitter – most importantly, the emitter must be extremely photostable, particularly at high laser excitation powers (tens of mW)[27]. Such a characteristic is rare, particularly at room temperature, where most applications of super-resolution nanoscopy are performed.

Here, we report two key results. First, we show that quantum emitters in hBN are sufficiently photostable to achieve super-resolution imaging of SPEs by GSD nanoscopy. Second, we harness the electronic structure and photophysical properties of these SPEs to demonstrate a new variant of GSD nanoscopy in which a low power (10 µW), doughnut-shaped, green laser is used in conjunction with the standard high power (tens of mW) excitation doughnut to achieve GSD imaging and realize a spatial resolution of ~(63±4) nm. We also demonstrate that the two-laser scheme can be used to reduce the excitation power needed to realize a given resolution target. This is relevant because the resolution of GSD nanoscopy scales with the excitation power and its principal limitation is the need for very high laser intensities, which may give rise to laser-induced damage for both the emitters and the surrounding materials[28]. Our results deepen the current understanding of the photophysics of quantum emitters in hBN and raise the interesting opportunity of extending their application to the ever-developing fields of super-resolution nanoscopy and bio-imaging.

A schematic of the hBN atomic lattice hosting the atomic defect is shown in Figure 1a, with blue and pink corresponding to the nitrogen and boron atoms, respectively. The atomic structure of the center is a matter of debate, and a number of vacancy-related defects have been proposed to be the origin of the quantum emission[9, 29, 30]. We first characterized the emitter using a conventional confocal, optical microscope and a Hanbury-Brown and Twiss (HBT) interferometer. Figure 1b shows the room-temperature photoluminescence (PL) spectrum of the emitter when excited with a 675-nm wavelength laser. The emitter has a zero phonon line (ZPL) at 778 nm and a negligible phonon sideband. The inset displays the second order correlation function, $g^{(2)}(\tau)$, which indicates that the emission is predominantly from a single defect: $g^{(2)}(\tau=0) \approx 0.25$, well below 0.5 at zero delay time (the correlation data are not background-corrected). Figure 1c shows the emitter saturation curve measured using a 708-nm laser as the excitation source; the 50% value of the saturated emission occurs at 14 mW. Emission polarization measurements of the emitter (figure 1c, inset) reveal that the emission is fully polarized, as expected from a dipole located in-plane within the layered host crystal.

After confirming the quantum nature of the emitter, we introduced a second laser to look for nonlinearities in the emission intensity. Figure 1d compares the emitter emission upon excitation with either a 675-nm (purple trace) or a 532-nm laser (green trace), with the emission produced under coincident excitation by both lasers (red trace). Note that the emitter was excited with only 10 µW of power for the 532-nm laser versus 300 µW for the 675-nm laser. The 532-nm excitation yields a negligible fluorescence intensity even when the corresponding PL spectrum shown in figure 1d (green trace) is multiplied tenfold. However, comparing the excitation of the same emitter with the 675-nm laser and co-excitation with the laser pair, i.e. 675-nm (300 µW) plus 532-nm (10 µW), reveals highly nonlinear behavior (purple and red traces in figure 1d). Upon co-excitation, the emission intensity increases by more than twofold, which is far greater than the 3.3% increase in total excitation power (from 300 to 310 µW). This behaviour, highlighted in figure 1e, is attributed to repumping of the emitter by one of the lasers (the 532-nm laser, as detailed below) which re-populates the excited bright state from metastable dark states (Supplementary information).

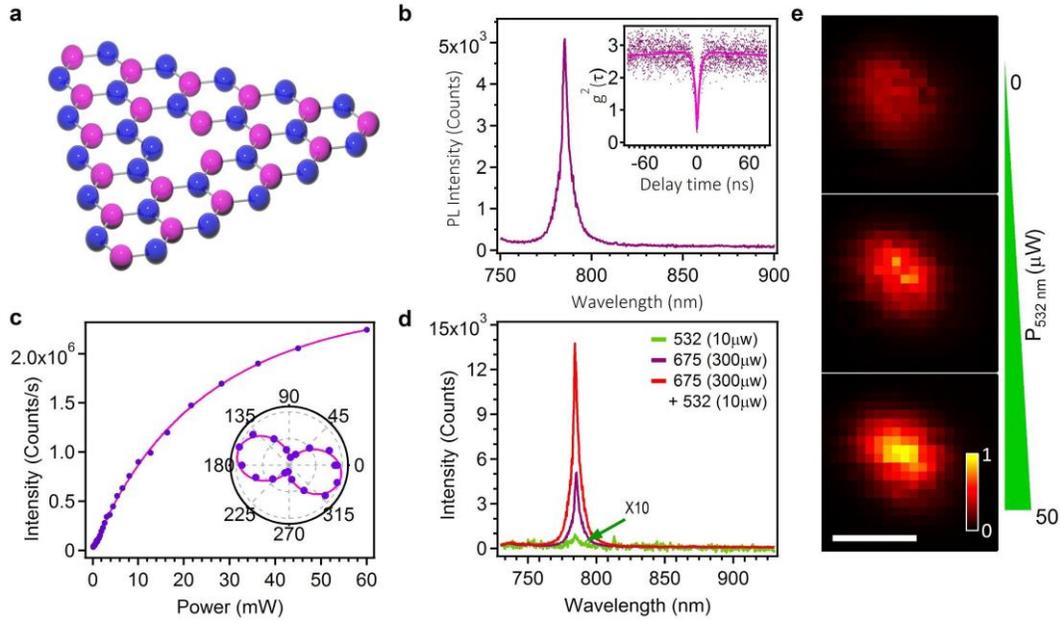

**Figure 1.** Single photon emission from hBN. **a)** Two-dimensional hBN hosting a fluorescent defect. **b)** Photoluminescence spectrum of the single defect in hBN under 675-nm excitation. Inset: second order autocorrelation measurement $g^{(2)}(\tau)$ of the defect with a dip at $g^{(2)}(0)$ of 0.25, indicating single-photon emission. **c)** Saturation curve of the emitter under excitation with a 708-nm laser. 50% of the saturated emission occurs at 14 mW. Inset: Emission polarization curve of the emitter. The emission dipole is in the plane of the layered host crystal. **d)** Photoluminescence spectra of the single defect under three excitation conditions: low-power (10 µW) 532-nm laser, high power 675-nm laser and coincident excitation with both lasers. Note that the spectrum under excitation with the 532-nm laser (green trace) has been multiplied tenfold for display purposes. **e)** Nonlinear increase of the emitter brightness upon a linear increase in the power of the 532-nm excitation laser coincident with the 675-nm excitation laser. Scale bar is 500 nm.

The emitter does not bleach even upon excitation by the highest laser powers (60 mW) used to obtain the saturation curve in figure 1e. Such photostability makes the emitter an appealing candidate for GSD nanoscopy. Before demonstrating GSD imaging, we first explore the details of the emitter photophysics which can be exploited in GSD nanoscopy. In particular, we focus on the rate kinetics of transitions to and from metastable dark states which can be manipulated optically to improve the resolution of GSD images.

Figure 2a shows the second order autocorrelation measurement recorded from the quantum emitter introduced in figure 1 under 675-nm laser excitation (pink curve), and upon co-excitation with the 675-nm laser and a variable power 532-nm laser (green arrow). The dip at short (ns) time scales confirms that the emitter is a single photon source with sub-Poissonian statistics. The exponential decays at longer (ms) time scales reveal the presence of additional metastable levels. The best fit to the data is achieved using a four-level model where two of the exponential decays correspond to metastable (dark) states (figure 2a). The corresponding time constants ($\tau_1$ and $\tau_2$) obtained from the fits are plotted in figure 2b (see methods). Interestingly, the two laser co-excitation measurements show that the metastable states are depopulated by the addition of the 532-nm laser – even at very low powers, ~0.1 µW. The time constants $\tau_1$ and $\tau_2$ plotted in figure 2b therefore decrease for increasing values of the 532-nm laser power (up to ~50 µW), in a nonlinear manner that correlates with the increase in brightness seen in figure 1d and 1e. Our interpretation is that adding the 532-nm laser repumps the system from the dark metastable states to the excited state from which the electron can recombine radiatively (more details in Supplementary information). The repumping also affects the saturation behavior of the emitter, as it acts to repopulate the short-lived excited state and deplete the long-lived metastable (dark) states. A simplified schematic of a level structure of the emitter that is consistent with these results is shown in figure 2c (the quantitative positions of the metastable states are not known). The repumping therefore reduces the power required to saturate the emitter, as shown in figure 2d where the saturation curve under 708-nm excitation (purple) is compared to those obtained upon co-excitation with the 532-nm laser at powers of 1, 5 and 10 µW. Under 708-nm excitation, the emitter saturates at ~14 mW. Conversely, during co-excitation, saturation is reached at ~3 mW ($P_{532nm}$ = 1 µW) and ~1.5 mW ($P_{532nm}$ = 5 µW or $P_{532nm}$ = 10 µW), respectively.

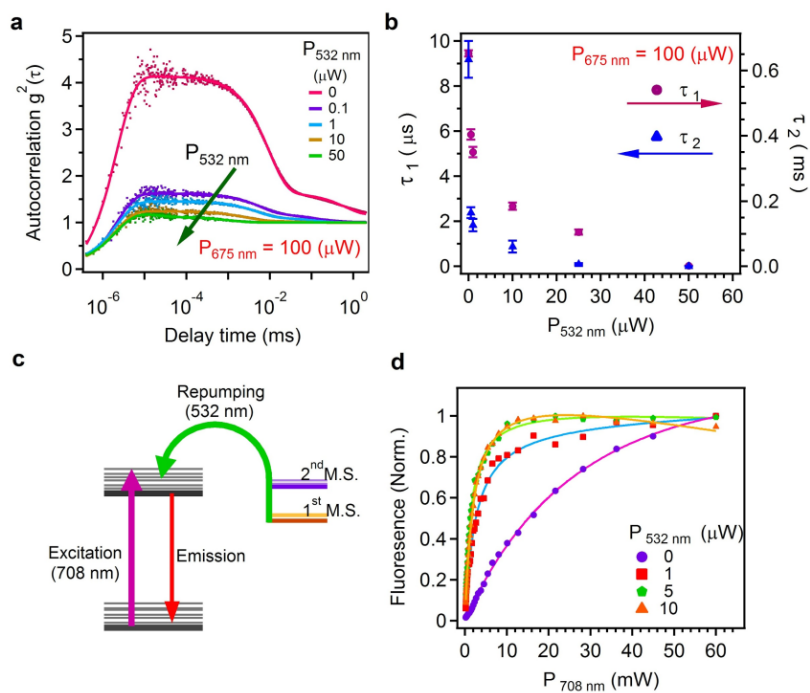

**Figure 2.** Photophysics of the single emitter introduced in figure 1. **a)** Long autocorrelation measurements under excitation with a 675-nm laser (100 µW). Increasing the power of a coincident laser (532-nm, green arrow) suppresses the population of the metastable states due to repumping of the emitter (see main

text). The experimental data (points) are fitted (solid lines) using a four-level model. **b)** Reduction in the time constants $\tau_1$ and $\tau_2$ associated with two metastable dark states (extracted from the fits in (a); see discussion in the main text) caused by increasing the power of the repumping 532-nm laser. **c)** Simplified level structure of the emitter showing the following transitions: excitation of the emitter from the ground state, radiative recombination that gives rise to the ZPL seen in figure 1d, and depopulation of metastable states by the 532-nm repumping laser based on the measurements in (a). M.S. = metastable state. **d)** The repumping causes a reduction in the laser power that is needed to saturate the emitter.

We now leverage the photophysical properties and repumping behavior of the emitter seen in figure 2 to perform and optimize the resolution of GSD nanoscopy. Figure 3a shows a schematic of our experimental setup. Briefly, we employed a vortex phase mask to modulate a 708-nm excitation laser and made it doughnut-shaped with an intensity that approached zero in the centre (Fig. 3a, inset). A second vortex phase mask was used to modulate a 532-nm laser, and the resulting doughnut-shaped intensity profile was co-aligned with the 708-nm doughnut in order to enable GSD imaging using a coincident laser pair. Both lasers were focused through an aberration-corrected objective (NA=0.95). The emitted photons were back-collected with the same objective, focused into the aperture of an optical fiber (used as a pinhole, similarly to a standard confocal microscope) and directed to an avalanche photodetector (see methods).

First, we performed negative GSD nanoscopy by using only the 708-nm doughnut-shaped laser as the excitation source (figure 3b). As the sample is scanned, the emitter experiences a doughnut-shaped excitation intensity profile which produces a corresponding 'high-null-high' emission pattern. In this configuration, the emitter location is given by the centre of the emission null. We can thus exploit the emitter saturation behaviour to achieve sub-diffraction resolution, as higher powers of the scanning doughnut beam generate steeper 'high-null' and 'null-high' PL emission gradients. This effectively narrows the full width at half maximum (FWHM) of the emission null, and the minimum in intensity yields an inverse image of the emitter with a spatial resolution that can exceed the diffraction limit – hence the denomination 'negative' GSD[26]. Deconvolution of the negative GSD image (left of figure 3b) yields a direct GSD image of the emitter (right of figure 3b). (see methods).

Figure 3d shows the GSD resolution that we can obtain by varying the power of the 708-nm doughnut-shaped laser, extracted from the FWHM of the null. With the experimental parameters of our setup (NA = 0.95, $\lambda_{Exc}$ = 708-nm), we achieve a resolution of (87±10) nm at 60 mW – well below ~460-nm, which is the diffraction-limited resolution of our confocal setup. Specifically, the resolution $\Delta r$ in GSD is given by[26]:

$$\Delta r \cong \lambda(\beta\pi n)^{-1}\sqrt{\epsilon + \frac{I_s}{I_m}} \qquad (1)$$

where $I_m$ is the maximum laser intensity in the crest of the doughnut, while $\epsilon I_m$ is the minimum ('null') intensity in the centre. The quantity $n$ is the refractive index of the medium, and $I_S$ is the laser intensity at which the emission intensity equals half of the maximum value in the limit of infinite excitation power. The parameter β is the 'steepness' of the point spread function (PSF) and depends both on the emitter properties and the 'crest-to-minimum' intensity gradient of the doughnut-shaped excitation source. In principle, GSD resolution is diffraction-unlimited and can be improved by increasing the excitation laser power beyond $I_s$ so as to minimize the ratio $I_s/I_m$. The consequent need to use high laser powers to achieve high spatial resolution is the main drawback of GSD and the related suite of RESOLFT (Reversible Saturable Optical Fluorescence Transitions) imaging techniques[28]. The high excitation powers needed for resolutions far beyond the diffraction limit cause bleaching of most emitters, and restrict the robust use of RESOLFT techniques to a limited number of systems such as highly stable color centres in diamond [31]. Furthermore, high excitation powers cause heating and damage of the surrounding environment and materials, which is particularly problematic for biological imaging applications of super-resolution nanoscopy.

The above problems caused by the need to use high laser powers can be alleviated if the photophysics of the emitter enables optical control of $I_S$. Specifically, if $I_S$ can be reduced then the resulting decrease in the ratio $I_s/I_m$ will improve GSD image resolution whilst maintaining a fixed laser power (and hence a fixed value of $I_m$ in equation 1). Such control is demonstrated by the PL saturation curves shown in figure 2c, where co-excitation with a low power 532-nm repumping laser is seen to cause a reduction in $I_S$. We therefore expect GSD resolution to improve if the imaging is performed using a coincident pair of excitation lasers. To verify this, we performed GSD imaging by adding a low-power, doughnut-shaped, 532-nm repumping laser co-aligned with the 708-nm doughnut beam. The addition of a 10 µW repumping beam indeed produces higher resolution images of the emitter, as is shown in figure 3c, where the corresponding direct images were again obtained via linear deconvolution. The improvement is illustrated directly in Figure 3e which compares the normalized intensity profile of the emitter excited by 40 mW of 708-nm beam before (violet circles) and after (orange triangles) applying the 10 µW re-pumping laser, as well as the PSF of our conventional confocal setup (green squares), obtained from the reflection image of a 50-nm gold nanoparticle.

A plot of GSD image resolution versus power, up to the maximum power achievable with our experimental setup, is shown in figure 3f for both the single (doughnut) excitation laser and the laser-doughnut pair. The highest resolution that we measured using the one and two-laser excitation schemes is (63±4) nm and (87±10) nm, respectively. Moreover, the power of the 708-nm laser needed to achieve a given target resolution is improved dramatically by the addition of the low power (10 µW) 532-nm repumping laser – for example, a resolution of 100 nm is achieved with powers of 55 mW and 30 mW when using the single and coincident laser excitation schemes, respectively. Such a reduction in the required laser power is highly desirable for super-resolution nanoscopy as it mitigates heating and damage caused by the high power excitation laser.

We note that prior implementations of dual-beam GSD imaging reported in the literature employ a coincident pair of doughnut and Gaussian-shaped laser beams[26, 27]. This configuration is inappropriate here since the low power 532-nm laser is used to reduce $I_S$ and this effect must be maximized in the crest of the doughnut-shaped high power beam whilst maintaining an intensity null at the beam axis (in the center of the doughnut).

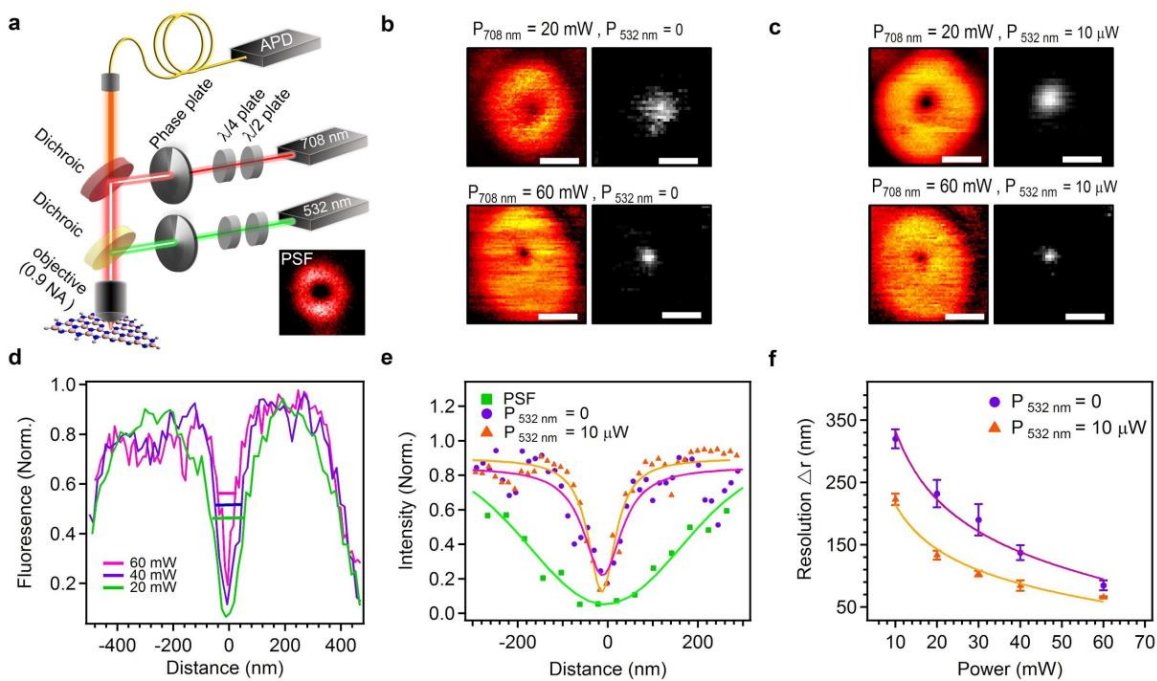

**Figure 3**. GSD nanoscopy of the emitter shown in figure 1 performed using a single laser and a coincident laser pair as the excitation source. **a)** Schematic of the setup used to perform GSD nanoscopy employing excitation lasers with doughnut-shaped intensity profiles. The system point spread function was determined by reflection of the lasers from a 50-nm gold nanosphere (inset, red circle). **b)** Negative GSD images of the single defect under excitation with a single 708-nm, doughnut-shaped laser, and using laser powers of 20 mW and 60 mW as indicated. The direct GSD images on the right are obtained by linear deconvolution of the negative GSD images on the left. **c)** Negative GSD images of the same emitter after addition of a 10 µw, 532-nm, doughnut-shaped re-pumping laser co-aligned with the 708-nm laser. The images on the right are linear deconvolutions of the negative ones. **d)** Photoluminescence intensity profiles of negative GSD in (b) showing a resolution improvement at higher excitation powers. Solid lines indicate the full width at half maximum (i.e. the resolution) of the emission null at the center of the doughnut. **e)** Intensity profiles used to compare the resolution obtained from negative GSD performed at 40 mW of 708 nm laser using the single doughnut beam (circles, violet) and the co-incident laser pair (triangles, orange). For comparison, the intensity profile obtained from reflection of 50 nm gold

nanoparticles (squares, green) is shown as point spread function of our setup. **f)** Dependence of GSD resolution on the power of the 708-nm laser, with and without the co-incident 532-nm, 10-µW repumping laser. Scale bars in (b) and (c) are 300 nm.

To conclude, we demonstrated super-resolution imaging of quantum emitters in a layered material – namely hBN. In particular, we performed GSD nanoscopy and achieved a resolution of ~(63±4) nm. Detailed photophysical analysis of the electronic level structure of the quantum emitters in hBN enabled us to develop a new modality of super-resolution microscopy that uses two coincident doughnut-shaped beams to improve resolution. We envision that this technique can be further adapted for other stable single emitters in 1D, 2D and 3D material hosts, as well as other fluorophores used for GSD nanoscopy. Our work advances the potential of defects in hBN as a promising candidate for bio-imaging applications as well as integrated quantum optics and nanophotonics.

**Acknowledgments**: The authors would like to acknowledge the Financial support from the Australian Research Council (DE130100592, DP140102721), FEI Company, the Asian Office of Aerospace Research and Development grant FA2386-15-1-4044. This research is supported in part by an Australian Government Research Training Program (RTP) Scholarship.

**Note**: During the preparation of the manuscript, we became aware of a related work from the group of A. Radenovic [ref 32] that exploits blinking behavior of localized emitters in hBN.

**Methods:**

**Sample preparation**. Graphene supermarket flakes were dropcast onto a silicon substrate and annealed at 850 C in Ar, in order to activate the emitters.

**Confocal and GSD microscopy**. Photoluminescence (PL) measurements were done in a home-built confocal setup. The sample was mounted onto a XYZ piezo stage (Physik Instrumente-Nanocube P-611) with positioning resolution of 0.2 nm. Excitation was performed using different laser sources: Ti:saph (M-squared, 700-750 nm), Supercontinuum (NKT photonics, Fianium WhiteLase supercontinuum laser) equipped with Acousto-optic Tunable Filter (AOTF, 400-550 nm ), 532-nm laser (Shanghi dreamlasers, 532nm low noise CW laser), 675-nm laser (PiL051XTM, Advanced Laser Diode Systems GmbH). To make a doughnut-shaped beam, the laser was first linearly polarized via a polarizing beam splitter (Thorlabs AR coated Cube beam splitter) and then passed through a zero-order half and a quarter phase plate (Thorlabs-Zero order waveplates) to achieve circular polarization. The beam was then directed through a vortex phase plate

(RPCphotonics, VPP-1b for 708 nm or VPP-1c for 532 nm lasers). The 708-nm and 532-nm lasers were guided to the sample using a long-pass filter (Semrock 785nm EdgeBasic) and a Dirchoic mirror (Semtock 532 nm dirchoic), respectively, and were focused on the sample through an aberration-corrected objective lens (Nikon 100X, NA = 0.95). The emission collected from the same objective was filtered using a notch filter (Semrock, 785 nm StopLin notch filter) and a 780-nm long-pass filter (Thorlabs, long pass color filter) and then coupled to the fiber which was connected to a spectrometer (Acton Spectra ProTM, Princeton Instrument Inc.) equipped with a 300 lines/mm grating and a charge-coupled device (CCD) detector with a resolution of 0.14 nm, or splitted into 50:50 in a Hanbury Brown and Twiss (HBT) interferometer for autocorrelation measurement using two avalanche photon detectors (Excelitas Technologies TM) and a time correlated counting module (Picoharp300TM, PicoQuantTM).

**Deconvolution**. To retrieve the direct image from the negative GSD scan, linear deconvolution was applied using built-in functions in Matlab. First, the high-resolution details (mainly the center local minimum in the negative GSD image) was removed using a short-pass Gaussian filter to produce a blurred image. Then the direct image was extracted by subtracting the GSD image from the blurred image. The blurred image from the application of the short-pass Gaussian filter (mathematical) method is preferred over the normal confocal scan of the image because of possible mismatch between confocal image and GSD image due to drifting during data acquisition. It is otherwise still possible to use the confocal map instead of the mathematical method.

**Autocorrelation data**. Autocorrelation data in figure 2a was fitted with the following equation:

$$g^2(\tau) = 1 - (1 + a_1 + a_2) \exp(\frac{\tau}{\tau_{exc}}) + a_1 \exp(\frac{\tau}{\tau_1}) + a_2 \exp(\frac{\tau}{\tau_2}) \quad (S1)$$

In the four-level model we used (see main text), three decay rates are considered for the excited state and two metastable dark states. The bunching time which corresponds to the excited state decay is shown in figure S5.